\documentstyle[preprint,prb,aps]{revtex}
\begin{document}
\draft
\title{Phenomenolgical model of elastic distortions near the
         spin-Peierls transition in $CuGeO_3$}
\author{M.L. Plumer}
\address{ Centre de Recherche en Physique du Solide et D\'epartement de
Physique}
\address{Universit\'e de Sherbrooke, Sherbrooke, Qu\'ebec, Canada J1K 2R1}
\date{June 1995}
\maketitle
\begin{abstract}
A phenomenological model of the Landau type forms the basis for a
study of elastic distortions near the spin-Peierls transition $T_c$
in $CuGeO_3$.  The atomic displacements proposed by Hirota {\it et al.}
[Phys. Rev. Lett. {\bf 73}, 736 (1994)] are accounted for by the model
which includes linear coupling between $Cu$ and $O$ distortions.
$Cu$ displacements are seen to be responsible for anomalies in the
elastic properties {\it at} $T_c$, whereas incipient $O$ distortions give
rise to temperature dependence below $T_c$.  A discussion of possible
critical behavior is also made.
\end{abstract}
%============================================================================
% BODY OF PAPER
\section{INTRODUCTION}
Due to the nature of its microscopic origins, attention has focused
on both magnetic\cite{mag} and structural
properties\cite{adams,nishi,lorenzo,arai,pouget,kamimura,hirota,roessli%
,harris,stpaul,nishi2,chen,poirier1,poirier2,poirier3,wink,yama}
of the recently discovered spin-Peierls system $CuGeO_3$.
It is now well established that this material is the first example
of an inorganic compound
which exhibits the simultaineous opening of a gap from a singlet
ground state and a lattice dimerization in S=1/2 spin chains.
In contrast with previously studied organic spin-Peierls systems,
large single crystals of $CuGeO_3$ can be made which facilitate
neutron and X-ray diffraction studies in addition to the measurments of
crystal-direction-dependent thermodynamic properties.
Such measurements have revealed large anisotropic
strain effects due to the dimerization, in contrast with simple
quasi-one-dimensional models of $Cu$ displacements.
Such effects are examined in the
present work based on a phenomenological Landau-type model which describes
the atomic displacements below the transition temperature $T_c$
as proposed by Hirota {\it et al.}\cite{hirota}
In contrast with the principal $Cu$ displacements which give rise to anomalies
{\it at} $T_c$,  secondary $O$ distortions are
found to be important in determining the temperature dependence of elastic
properties below $T_c$.   A linear coupling between the two
types of ionic displacements is proposed which is electronic in origin.
Good agreement with available experimental results is demonstrated.
This study also serves to compliment and extend the brief Landau-type
descriptions of some elastic properties given in Refs.\onlinecite{nishi,harris%
,poirier2}.

Of the ten ions per high-temperature (above $T_c$) orthorhombic
unit cell ($a$ = 4.81, $b$ = 8.47, and $c$ = 2.941 $\AA$),
at sites ${\bf R} = {\bf R}_l + {\bf v}_j$ (where
${\bf R}_l$ are Bravais lattice vectors and $j=1-10$), there are
two $Cu$ and four $O$ ions which displace at $T_c$ in the model of
Ref\onlinecite{hirota}.  These are shown schematically in Fig. 1, with the
high-T positional parameters ${\bf v}_j$ given by
\begin{eqnarray}
Cu_1:~v_1 &=&(0,0,0);~~~~~~~~~~Cu_2:~v_2 = (0,{\textstyle \frac12},0)
                    \nonumber \\
O_3: ~v_3 &=&(x,-y,{\textstyle \frac12});
               ~~~~~~~~O_4: ~v_4 = (-x,y,{\textstyle \frac12}) \\  \nonumber
O_5: ~v_5 &=&(-x,{\textstyle \frac12}-y,{\textstyle \frac12});
             ~~~O_6:~v_6 = (x,y+{\textstyle \frac12},{\textstyle \frac12}).
\end{eqnarray}
These displacements thus double the unit cell in the $a$ and $c$ directions
below $T_c$.  A simple order-parameter description of the ionic movements
below $T_c$ may be written as
\begin{equation}
\mbox{\boldmath $\rho(R)$} = \mbox{\boldmath $\rho$}_j \cos ({\bf Q \cdot R}_l)
\end{equation}
where ${\bf Q} = (\frac12,1,{\textstyle \frac12})$ and here $j=1-6$.
The displacements shown in Fig. 1 are then described with
\begin{eqnarray}
\mbox{\boldmath $\rho$}_1 = - \rho_z {\bf \hat z};
         ~~\mbox{\boldmath $\rho$}_2 = - \mbox{\boldmath $\rho$}_1 \nonumber \\
\mbox{\boldmath $\rho$}_3 = - \rho_x {\bf \hat x} - \rho_y {\bf \hat y};
       ~~\mbox{\boldmath $\rho$}_4 = - \mbox{\boldmath $\rho$}_3 \\  \nonumber
\mbox{\boldmath $\rho$}_5 = - \rho_x {\bf \hat x} + \rho_y {\bf \hat y};
          ~~\mbox{\boldmath $\rho$}_6 = - \mbox{\boldmath $\rho$}_5
\end{eqnarray}
where $\rho_x$ and $\rho_y$ are associated with the $O$ distortions
and $\rho_z$ represents $Cu$-ion displacements.  It is argued below that
simple ion-ion interactions derived from symmetry arguments can
reproduce the configuration of displacements depicted in Fig. 1.  A
desirable feature of any such model is to account for the simultaneous
displacement of both $Cu$ and $O$ ions (which is also relevant to possible
critical behavior).  It is for this reason that
that ionic density-density coupling must also be considered.  In addition,
simple coupling between $\mbox{\boldmath $\rho$}$
and the strain tensor reproduces the principal
features of measured elastic properties without evoking higher-order
interactions as in Refs.\onlinecite{nishi,harris,poirier2}.

\section{LANDAU FREE ENERGY}

Landau-type free energies may be constructed using symmetry
arguments\cite{landau}.  It is assumed here that magnetic degrees
of freedom are integrated out so that the free energy
depends explicitly on \mbox{\boldmath $\rho$} only.
In the case of orthorhombic symmetry, there
are three independent invariants which contribute to free energy density
$f$ at second order in the vector \mbox{\boldmath $\rho$},
which may be chosen as
\begin{equation}
f_2 = D'_x \rho_x^2 + D'_y \rho_y^2 +D'_z \rho_z^2.
\end{equation}
The independence of these three coefficients requires that
$\rho_x \not= \rho_y \not= \rho_z$, as observed experimentally.
Alternatively, one can choose the three terms as
$f_2 = A \rho^2 + D_x \rho_x^2 +D_y \rho_y^2$,
where the $A$-term represents an isotropic interaction.
Minimization of the anisotropy at the $Cu$ sites, for
example, gives
$\mbox{\boldmath $\rho$}_1, \mbox{\boldmath $\rho$}_2
                                      \parallel \pm {\bf \hat z}$
if $D_x(Cu) > 0$ and $D_y(Cu) > 0$.
In the following
analysis, only isotropic terms are considered in detail, which is
sufficient for the purposes of the present study.

\subsection{Second-order terms}

A general expression for isotropic two-ion interactions is
\begin{equation}
F_2 = \frac12 \sum_{{\bf R,R'}} A{\bf(R - R')}
     \mbox{\boldmath $\rho$}({\bf R}) \cdot \mbox{\boldmath $\rho$}({\bf R'}).
\end{equation}
Using the form (2) for $\mbox{\boldmath $\rho$({\bf R})}$
in this relation yields
\begin{equation}
F_2 = \frac12 \sum_{j,k} {\tilde A}_{jk}({\bf Q})
         \mbox{\boldmath $\rho$}_j \cdot \mbox{\boldmath $\rho$}_k,
\end{equation}
where
\begin{equation}
{\tilde A}_{jk}({\bf Q}) = \sum_{\mbox{\boldmath $\tau$}}
                A(\mbox{\boldmath $\tau$} + {\bf v}_j - {\bf v}_k)
                \cos ({\bf Q} \cdot \mbox{\boldmath $\tau$})
\end{equation}
with $\mbox{\boldmath $\tau$} = {\bf R}_l - {\bf R'}_l$.
Considered below are the near-neighbor interactions within and between
the $CuO_2$ chains which are likely to be important.  We recall
that $CuGeO_3$ is somewhat quasi-one-dimensional.

\subsubsection{Cu-Cu within chains}

Interactions between near-neighbor $Cu$ ions along the $c$ axis
involve $j=k=(1,2)$ with ${\tilde A}_{11} = A({\bf c}) \cos \pi
+ A({-\bf c}) \cos \pi$,
and with a similar expression for ${\tilde A}_{22}$, so that
\begin{eqnarray}
F_{2A} &=& -A({\bf c})[\mbox{\boldmath $\rho$}_1 \cdot \mbox{\boldmath
$\rho$}_1
 + \mbox{\boldmath $\rho$}_2 \cdot \mbox{\boldmath $\rho$}_2 ] \nonumber \\
       &=& -2A_{11} \rho_C^2
\end{eqnarray}
where $A_{11} = A({\bf c})$.

\subsubsection{Cu-Cu between chains}

Here $j \not= k=(1,2)$ with, e.g.,
${\tilde A}_{12} = A(-{\bf v}_2) + A({-\bf v}_2 + b{\bf \hat y}) \cos2 \pi$,
giving
\begin{equation}
F_{2B} = 2A({\bf v}_2)
\mbox{\boldmath $\rho$}_1 \cdot \mbox{\boldmath $\rho$}_2.
\end{equation}
If $A({\bf v}_2) > 0$, then this term is minimized by the configuration
$\mbox{\boldmath $\rho$}_1 \parallel -\mbox{\boldmath $\rho$}_2$, as in
Fig. 1.  It can be expected, however, that this term is smaller
than $F_{2A}$ due to the larger interaction distance involved.
Its contribution is given by $F_{2B} = -2A_{12} \rho_C^2$,
where $A_{12} = A({\bf v}_2)$.

\subsubsection{O-O within chains}

Interactions between the two $O$ ions which surround the $Cu$ of Fig. 1
correspond with $j \not= k=(3,4)$ and $(5,6)$.  Using
${\tilde A}_{34} = A({\bf v}_3 - {\bf v}_4) = A_{43} = A_{56} = A_{65}$,
the energy is
\begin{equation}
F_{2C} = A({\bf v}_3 - {\bf v}_4)
[\mbox{\boldmath $\rho$}_3 \cdot \mbox{\boldmath $\rho$}_4
+ \mbox{\boldmath $\rho$}_5 \cdot \mbox{\boldmath $\rho$}_6].
\end{equation}
If $A({\bf v}_3 - {\bf v}_4) > 0$, then the configuration
$\mbox{\boldmath $\rho$}_3 \parallel -\mbox{\boldmath $\rho$}_4$ and
$\mbox{\boldmath $\rho$}_5 \parallel -\mbox{\boldmath $\rho$}_6$,
as in Fig. 1, is stabilized by this term, giving
$F_{2C} = -2A_{34} \rho_O^2$, where $A_{34} = A({\bf v}_3 - {\bf v}_4)$.

\subsubsection{O-O between chains}

With $j \not= k=(4,5)$ and $(3,6)$, the resulting contribution to the
free energy is
\begin{equation}
F_{2D} = A({\bf v}_4 - {\bf v}_5)
[\mbox{\boldmath $\rho$}_4 \cdot \mbox{\boldmath $\rho$}_5
+ \mbox{\boldmath $\rho$}_3 \cdot \mbox{\boldmath $\rho$}_6].
\end{equation}
Although parallel or anti-parallel configurations are minimized
by this term, it is expected to be rather small.  Fourth-order
terms considered below prefer a configuration
$\mbox{\boldmath $\rho$}_4 \perp \mbox{\boldmath $\rho$}_5$ and
$\mbox{\boldmath $\rho$}_3 \perp \mbox{\boldmath $\rho$}_6$
giving $F_{2D} = 0$.  As emphasized above, however, anisotropy effects
will alter somewhat this idealized state.

\subsubsection{Cu-O within chains}

Considered here are the near-neighbor $Cu-O$ interactions, so that
$j \not= k=(1,3)$, (1,4), (2,5), as well as (2,6).  Due to the fact
that neighboring $Cu$ ions along the $c$ axis displace in opposite
directions, and that the $O$ ions (3,4) and (5,6) are
located mid-way between these $Cu$ ions, this two-ion interaction term
is zero.  This is illustrated by considering
${\tilde A}_{13} = A(-{\bf v}_3) + A({\bf c} - {\bf v}_3) \cos \pi = 0$,
where $A(-{\bf v}_3) = A({\bf v}_3) = A({\bf c} - {\bf v}_3)$ has
been used.  In the absence of $Cu-O$ coupling at second order, there
is not a simultaneous displacement of both species.  This result is
not in agreement with the observed behavior.

\subsubsection{Cu-O within chains: density-density interactions}

In addition to the two-ion intereactions which involve the vector
displacements $\mbox{\boldmath $\rho(R)$}$, there are also ionic
charge density couplings which involve the perturbed density
$\rho({\bf R})~=~|\mbox{\boldmath $\rho(R)$}|$.
Such interactions are of the type
which have been considered within the framework of density functional
theory\cite{plum} and are of the form
\begin{eqnarray}
F_G &=& \frac12 \sum_{{\bf R,R'}} G{\bf( R -  R')}
      \rho({\bf R}) \rho({\bf R'}) \nonumber \\
    &=& \frac12 \sum_{j,k} {\tilde G}_{jk} \rho_j \rho_k
\end{eqnarray}
where
\begin{equation}
{\tilde G}_{jk} = \sum_{\mbox{\boldmath $\tau$}}
                G(\mbox{\boldmath $\tau$} + {\bf v}_j - {\bf v}_k).
\end{equation}
For simplicity, only the near-neighbor $Cu-O$ coupling is explicitly
considered here (the other interactions discussed above are simply
renormalized by charge density coupling).  This gives a contribution
\begin{equation}
F_G = 4 G_{13} \rho_C \rho_O,
\end{equation}
where $G_{13} = G({\bf v}_3)$.
Linear coupling of this type yields (see below)
$\rho_O \propto \rho_C$, as expected.

Temperature dependence appears through the usual entropy
term of mean-field theory\cite{landau}
\begin{equation}
F_e = {\textstyle \frac12} aT \sum_{\bf R} \rho^2({\bf R}).
\end{equation}
With the simplifying assumption that $\rho_x \simeq \rho_y$ (see below),
all of the second-order terms considered above
can be written as
\begin{equation}
F_2 = A_C \rho^2_C + A_O \rho^2_O + A_{CO} \rho_C \rho_O,
\end{equation}
where
\begin{equation}
A_C = aT - 2A_{11} - 2A_{12}, ~~~ A_O = aT - 2A_{34}, ~~~ A_{CO} = 4G_{13}.
\end{equation}

\subsection{Fourth-order terms}

We omit here consideration of fourth-order anisotropy terms and note that,
in general, isotropic interactions involving the vector
\mbox{\boldmath $\rho(R)$} are nonlocal and can be expressed as\cite{plum2}
\begin{equation}
F_4 = \frac14 \sum_{\lbrace{\bf R}_i \rbrace} B{\bf(R_1,R_2;R_3,R_4)}
[\mbox{\boldmath $\rho$({\bf $R_1$})}
               \cdot \mbox{\boldmath $\rho$({\bf $R_2$})}]
[\mbox{\boldmath $\rho$({\bf $R_3$})}
              \cdot \mbox{\boldmath $\rho$({\bf $R_4$})}],
\end{equation}
where $B$ depends only on the relative distances $\mid {\bf R_i - R_j} \mid$.
For simplicity, and for the puposes of demonstrating that the present model
is plausible, only two types of terms will be considered explicitly here:
\begin{eqnarray}
F_{4a} &=& \frac14 B_0 \sum_{\bf R}
  [\mbox{\boldmath $\rho$({\bf R})}
        \cdot \mbox{\boldmath $\rho$({\bf R})}]^2 \nonumber \\
       &=& B_0[ \rho_C^2 + 2 \rho_O^2 ]^2,
\end{eqnarray}
which comes from entropy considerations,\cite{landau}
as well as the simple two-ion interactions of the form
\begin{equation}
F_{4b} = \frac14 \sum_{{\bf R,R'}} B{\bf(R -  R')}
  [\mbox{\boldmath $\rho$({\bf R})} \cdot \mbox{\boldmath $\rho$}({\bf R'})]^2.
\end{equation}
The five near-neighbor couplings as considered above for $F_2$ yield a result
\begin{eqnarray}
F_{4b} &=& B_{11} [\rho_1^4 + \rho_2^4]
 + B_{12}(\mbox{\boldmath $\rho$}_1
          \cdot \mbox{\boldmath $\rho$}_2)^2 \nonumber \\
       &+& {\textstyle \frac12} B_{34}[(\mbox{\boldmath $\rho$}_3
                          \cdot \mbox{\boldmath $\rho$}_4)^2
                   +  (\mbox{\boldmath $\rho$}_5
                          \cdot \mbox{\boldmath $\rho$}_6)^2]
+ {\textstyle \frac12} B_{45}[(\mbox{\boldmath $\rho$}_4
                          \cdot \mbox{\boldmath $\rho$}_5)^2
                   + (\mbox{\boldmath $\rho$}_3
                         \cdot \mbox{\boldmath $\rho$}_6)^2]      \nonumber \\
  &+& B_{13}[(\mbox{\boldmath $\rho$}_1 \cdot \mbox{\boldmath $\rho$}_3)^2
          + (\mbox{\boldmath $\rho$}_1 \cdot \mbox{\boldmath $\rho$}_4)^2
          + (\mbox{\boldmath $\rho$}_2 \cdot \mbox{\boldmath $\rho$}_5)^2
          + (\mbox{\boldmath $\rho$}_2 \cdot \mbox{\boldmath $\rho$}_6)^2].
\end{eqnarray}
Configurations minimized by these terms are as follows:
\begin{eqnarray}
For ~ B_{12} & < & 0, ~~~
 \mbox{\boldmath $\rho$}_1 \parallel \pm \mbox{\boldmath $\rho$}_2; \nonumber
\\
for ~ B_{34} & < & 0, ~~~
 \mbox{\boldmath $\rho$}_3 \parallel \pm \mbox{\boldmath $\rho$}_4; ~~
 \mbox{\boldmath $\rho$}_5 \parallel
            \pm \mbox{\boldmath $\rho$}_6;\nonumber \\
for ~ B_{45} & > & 0, ~~~
           \mbox{\boldmath $\rho$}_4 \perp  \mbox{\boldmath $\rho$}_5; ~~
 \mbox{\boldmath $\rho$}_3 \perp  \mbox{\boldmath $\rho$}_6;\nonumber \\
for ~ B_{13} & > & 0, ~~~
           \mbox{\boldmath $\rho$}_C \perp  \mbox{\boldmath $\rho$}_O.
\end{eqnarray}
Note that the complete minimization of the term $B_{45}$ occurs only if
$\rho_x = \rho_y$ when (3) is used, but that
this equality is in competition with anisotropy terms.  It
appears, however, that this condition is nearly satisfied in $CuGeO_3$.
These results, together with the minimization conditions from consideration
of $F_2$, fully explain the displacements of Fig. 1, as described by Eqs.
(1)-(3).

\subsection{Equilibrium Properties}

The full free energy is then given by
\begin{equation}
F = A_C \rho^2_C + A_O \rho^2_O + A_{CO} \rho_C \rho_O
 + B_C \rho^4_C + B_O \rho^4_O + B_{CO} \rho_C^2 \rho_O^2,
\end{equation}
where $B_C = B_0 + B_{11} + B_{12}$, $B_O = 4B_0 + B_{34}$, and $B_{CO} =
4B_0$.
Due to the nature of the spin-Peierls interaction, it is assumed here that
$Cu$ and $O$ displacements represent primary and secondary order parameters,
respectively.
Minimization of $F$ with respect to $\rho_O$ yields a result which can
be expressed as an expansion in odd powers of $\rho_C$, at least near the
transition temperature,
\begin{equation}
\rho_O  = \alpha \rho_C + \beta \rho^3_C + \cdot \cdot \cdot,
\end{equation}
where
\begin{equation}
\alpha = - {\textstyle \frac12} A_{CO} / A_O; ~~~
\beta = - \alpha / A_{O} (B_{CO} + {\textstyle \frac12} B_O A_{CO}^2 / A_O^2).
\end{equation}
This establishes a linear relationship between $Cu$ and $O$ displacements
near $T_c$, but that a more complicated behavior can be expected at lower
temperatures.  Using this result in the above expression for $F$ then
gives simply
\begin{equation}
F = A \rho^2_C + {\textstyle \frac12} B \rho^4_C,
\end{equation}
where $A = A_C - \frac12 \alpha A_{CO}$ and
$\frac12 B = B_C + \alpha^2 B_{CO} + \alpha^4 B_O$.
The transition temperature is defined by $A \equiv a(T - T_c)$, so that
\begin{equation}
aT_c = A_+  +  \left \lbrace A_-^2 + 4G_{13}^2 \right \rbrace ^\frac12,
\end{equation}
where $A_\pm = A_{11} + A_{12} \pm A_{34}$ and $A_{CO} = 4 G_{13}$ has been
used.

For later purposes, we note here that minimization of $F$ above yields
$\rho_C^2 = -A/B = (a/B)(T_c-T)$ for the primary order parameter,
and that the specific-heat anomaly at $T_c$ is given by
\begin{equation}
\Delta C = C(T_c^+) -  C(T_c^-) = -T_c \frac{a^2}{B}.
\end{equation}

\section{ELASTIC COUPLING}

The orthorhombic symmetry of $CuGeO_3$ allows for nine independent
elastic constants: $C_{11}, C_{12}, C_{13}, C_{22}, C_{23}, C_{33},
C_{44}, C_{55}, C_{66}$.  The elastic energy is conveniently
written as
\begin{equation}
F_{el} = \frac12 \sum_{\mu \nu} C_{\mu \nu} e_\mu e_\nu
\end{equation}
where Voigt notation $(1 \equiv xx, 2 \equiv yy, 3 \equiv zz,
4 \equiv yz, 5 \equiv zx, 6 \equiv xy)$ has been used.

Lowest order coupling between the strain tensor and displacement
vector \mbox{\boldmath $\rho$} is of the form $ \sim e \rho^2$
and also has nine independent terms
\begin{eqnarray}
F_{\rho e} = \frac12 \sum_{{\bf R,R'}} \lbrack &D_1& e_{xx} \rho_x \rho_x
+ D_2 e_{yy} \rho_y \rho_y + D_3 e_{zz} \rho_z \rho_z
+ D_4 (e_{xx} \rho_y \rho_y + e_{yy} \rho_x \rho_x) \nonumber \\
+ &D_5& (e_{xx} \rho_z \rho_z + e_{zz} \rho_x \rho_x)
+ D_6 (e_{yy} \rho_z \rho_z + e_{zz} \rho_y \rho_y)  \\
+ &D_7& e_{yz} \rho_y \rho_z  + D_8 e_{xz} \rho_x \rho_z
+ D_9 e_{xy} \rho_x \rho_y   \rbrack,                \nonumber
\end{eqnarray}
where the short-hand notation $D_i = D_i({\mbox{\boldmath $\tau$}})$ and
$\rho_\alpha \rho_\beta = \rho_\alpha({\bf R}) \rho_\beta({\bf R'})$
has been used.  The same near-neighbor interactions as considered
above for $F_2$ and $F_{4b}$ yield
\begin{eqnarray}
F_{\rho e} = 2 \lbrack &D_1& e_{xx} \rho_x^2
+ D_2 e_{yy} \rho_y^2 + D_3 e_{zz} \rho_z^2
+ D_4 (e_{xx} \rho_y^2 + e_{yy} \rho_x^2)\nonumber \\
+ &D_{5C}& e_{xx} \rho_z^2 + D_{5O} e_{zz} \rho_x^2
+ D_{6C} e_{yy} \rho_z^2 + D_{6O} e_{zz} \rho_y^2
+ D_9 e_{xy} \rho_x \rho_y   \rbrack.
\end{eqnarray}
Recall that $Cu$ displacements are along the $z$ axis and that
$O$ displacements are in the $xy$ plane.  The interaction constants
$D_1, D_2, D_4, D_{5O}, D_{6O}$, and $D_9$ thus involve $O-O$ couplings
whereas $D_3, D_{5C}$, and $D_{6C}$ involve $Cu-Cu$ couplings.
$Cu-O$ interactions are zero for the same reasons as described in section
2A-5, although those inlvolving density-density couplings could be
added if desired.  With the assumption $\rho_x =\rho_y = \rho_O/ \sqrt{2}$,
as well as the result from Eq. (24)
\begin{equation}
\rho_O^2 = \chi \rho_C^2, ~~~ \chi \simeq \alpha^2 + 2 \alpha \beta \rho_C^2,
\end{equation}
the elastic coupling term reduces to
\begin{eqnarray}
F_{\rho e} &=&  \lbrack K_1 e_{xx} +K_2 e_{yy} + K_3 e_{zz}
+ K_6 e_{xy} \rbrack \rho_C^2 \nonumber \\
           &=& \sum_\mu  K_\mu e_\mu \rho_C^2,
\end{eqnarray}
where
\begin{eqnarray}
K_1 &=& (D_1 + D_4) \chi + 2 D_{5C} \nonumber \\
K_2 &=& (D_2 + D_4) \chi + 2 D_{6C} \nonumber \\
K_3 &=& 2 D_3 (D_{5O} + D_{6O}) \chi \nonumber \\
K_6 &=& D_9 \chi.
\end{eqnarray}
Note that $\chi$ is temperature dependent through $\rho_C^2$ and that it
is proportional to the square of the density-density coupling constant,
$\chi \sim G_{13}^2$.  It is also convenient to write
\begin{equation}
K_\mu = K_{\mu o} + k_\mu \rho_C^2,
\end{equation}
where $K_{\mu o}$ is independent of temperature and depends
only on $Cu-Cu$ interactions, whereas the coefficient of the
temperature-dependent contribution $k_\mu$ depends only on $O-O$ interactions.

\subsection{Thermal expansion and elastic constants}

The full free energy now takes the simple form\cite{plum2}
\begin{equation}
F = A \rho^2_C + {\textstyle \frac12} B \rho^4_C
+ {\textstyle \frac12} \sum_{\mu \nu} C_{\mu \nu} e_\mu e_\nu
+  \sum_\mu  K_\mu e_\mu \rho_C^2.
\end{equation}
Minimization with respect to the strain tensor yields the following
results for the effects of $Cu$ and $O$ displacements on
thermal expansion, the expansion coefficient, and the elastic constants
\begin{eqnarray}
e_\mu &=& -\sum_\nu s_{\mu \nu} \lbrack K_{\nu o}
            + k_\nu \rho_C^2 \rbrack \rho_C^2  \nonumber \\
\Delta \alpha_\mu &=& - (a/B) \sum_\nu s_{\mu \nu} \lbrack K_{\nu o}
+ 2 k_\nu \rho_C^2  \rbrack \nonumber \\
\Delta C_{\mu \nu} &=& (K_{\nu o} + k_\nu \rho_C^2)
                            (K_{\mu o} + k_\mu \rho_C^2) /B
\end{eqnarray}
where $s_{\mu \nu}$ is the compliance matrix
$(s_{\mu \nu} = C^{-1}_{\mu \nu})$,
$\Delta \alpha_\mu = \alpha_\mu (T_c^+) - \alpha_\mu (T_c^-)$,
and with a similar definition for $\Delta C_{\mu \nu}$.
Due to the temperature dependence of $\rho_C^2$, thermal
expansion is not simply proportional to the square of the
order parameter, and the expansion coefficient does not exhibit a simple
step discontinuity, as is usually the case with low order
($ \sim e \rho^2$) interactions.  In addition, the elastic constants exhibit
temperature dependence below $T_c$, also due to the displacement
of $O$ ions (since $k_\mu \rho_C^2 \sim \rho_O^2$).
Such effects can also be mimicked by including higher-order
couplings.\cite{nishi,harris,poirier2}

\subsection{Numerical fits}

With available experimental data, it is possible to fit a number
of the Landau model parameters.
{}From the specific heat data of Ref.\onlinecite{wink},
the estimate $\Delta C \simeq -4.5 \times 10^4 J/(K m^3)$ can be made.
Comparing this result with Eq. (28), and using $T_c = 14.2K$, gives
$a^2/B \simeq 3.2 \times 10^3 J/m^3$ (where the free energy $F$ is in units
of energy/volume).  The order parameter at low temperature takes the form
$\rho_C(0) = (a T_c/B)^\frac12$.  This expression can then be used with
the above result, and the estimate from Hirota {\it et al.} of
$\rho_C(0) \simeq 0.0014 c$, to give
$a \simeq 2.6 \times 10^{29} J/(K m^5)$ and
$B \simeq 2.2 \times 10^{55} J/(K m^7)$ in (cumbersome) $SI$ units.

There are also available limited data on the elastic constants near
$T_c$.  In particular, from the results of Ref.\onlinecite{poirier1}
on $C_{11}, C_{22}$, and $C_{33}$, it is clear that only $C_{33}$
exhibits a detectable discontinuity so that $K_{10} \simeq 0$,
$K_{20} \simeq 0$.  From these data, the estimate
$\Delta C_{33}/C_{33} \simeq 2.5 \times 10^{-3}$ can be made.  Using
this result, along with\cite{private} $C_{33} \simeq 3.7 \times 10^{11} N/m^2$,
as well as Eq. (37) and the above value for $B$, the estimate
$K_{30} \simeq 1.5 \times 10^{32}J/m^5$
can be made.  These results, together with the relations (37),
imply that the only other elastic constants which could display a
discontinuity at $T_c$ are $C_{36}$ and $C_{66}$, provided that $K_{60}$ has
appreciable magnitude.

The temperature dependence of $\Delta C_{\mu \nu}$ below $T_c$ from Eq. (37)
is of the form $\sim  \Delta C_0 + c_1 \rho^2_C + c_2 \rho^4_C$,
where in general $\rho_C \sim (T_c - T)^\beta$.  From the data of
Ref.\onlinecite{poirier1}, there appears to be a region near $T_c$ where
$\Delta C_{11}$, $\Delta C_{22}$, and $\Delta C_{33}$ increase
{\it nearly} linearly with decreasing temperature,
followed by a less rapid increase at lower temperatures.  A more detailed
analysis of these results\cite{poirier2}
suggests a value $\beta = 0.42(2)$.  This estimate may be
compared with $\beta = 0.25(5)$ found in an earlier study
of elastic properties.\cite{stpaul} A discussion of
possible critical behaviors is deferred to the next section.

It is also of interest to examine the expression for the effective
fourth-order coefficient $B' = B - \sum_{\mu \nu} K_{\mu 0} K_{\nu 0}$
which results from elastic coupling.\cite{plum2}  With the reasonable
assumptions that the dominant contribution is from $K_{30}$, and
that $s_{33} \sim 1/C_{33}$, one finds $(B' - B)/B \sim 10^{-3}$.  The
argument given in Ref.\onlinecite{harris} that elastic coupling
could be responsible for reducing the magnitude of the fourth-order
coefficient to near zero appears to be inconsistent with the above
analysis.  Thus, the proposal in Ref.\onlinecite{harris}
of tricriticality is not supported by the present work.

Experimental results on the three diagonal components of the
strain tensor may be found in Ref.\onlinecite{wink}.  From Eq. (37)
and the approximations $K_{10} = K_{20} = 0$,
temperature dependence of the form
\begin{equation}
e_\mu =  -\lbrack s_{\mu 3} K_{30} + (s_{\mu 1} k_1 + s_{\mu 2} k_2)
+ s_{\mu 3} k_3) \rho^2_C  \rbrack \rho^2_C ,
\end{equation}
is expected.
It is clear that such predicted behavior is consistent with the
data.  Detailed fitting to extract estimates of the three parameters
$k_\mu$ must await determination of the absolute values of
the other elastic constants (only $C_{11}, C_{22}$, and $C_{33}$ are
presently available) and hence knowledge of the compliance matrix.
With the assumption of quadratic order-parameter dependence, the authors
of Ref.\onlinecite{wink} extracted estimates $2 \beta = 0.61(2)-0.69(8)$.

Discontinuities in $\Delta \alpha_\mu$ at $T_c$ along the three principle
axes $a, b$, and $c$ were found to be $1.3(1)$, $-2.3(1)$,
and $-0.4(1) \times 10^{-5}/K$, respectively.\cite{wink}
{}From (37), as well as the approximation $s_{33} \sim 1/C_{33}$, the
c-axis expansivity is given by $\Delta \alpha_c \simeq -(a/B)K_{30}/C_{33}$.
Using the parameter estimates given above, this yields the model
prediction $\Delta \alpha_c \simeq  -5 \times 10^{-6}$, in very good
agreement with the data.  Comparison between the model predictions
$\Delta \alpha_a = -(a/B)K_{30}s_{13}$ and
$\Delta \alpha_b = -(a/B)K_{30}s_{23}$ at $T_c$
with corresponding experimental results
requires knowledge of the compliance matrix.
It is also of interest to note that the temperature dependences observed
below $T_c$ in the expansivity are accounted for by Eq. (37).

\section{CONCLUSIONS}

In spite of the great interest in the novel properties of the new
inorganic spin-Peierls system $CuGeO_3$ as reflected by the large number
of recently published experimental studies, there has been relatively
few theoretical investigations.  This is partly due to the complex
nature of the relevant interactions caused by the non-negligible
interchain interactions in this not-so quasi-one-dimensional system.
Phenomenological models are thus particularly useful due to their
simplicity and serve to guide further study of a more microscopic nature.
The Landau-type free energy studied in the present work was constructed
from symmetry arguments with dominant close-neighbor interactions
included.  The analysis demonstrates that the simplest of interactions can
account for the model of rather complicated ionic displacements proposed in
Ref.\onlinecite{hirota}.  $O$-ion distortions are coupled linearly to
the principal $Cu$-ion displacements by charge-density interactions.
Elastic distortions as a function of temperature below $T_c$, not
present in simple Landau models, are seen here to be a consequence of
this incipient $O$-ion movement.  This contrasts with the elastic anomalies
{\it at} $T_c$, which are due to the $Cu$ displacements.

A simple argument based on the results of this work suggest that the
transition should belong to the standard $XY$ universality class
(where, e.g., $\alpha = -0.012(10), \beta=0.349(4), \nu = 0.671(5)$).
One component is simply the $Cu$ displacement along the $c$ axis,
$\rho_z = \rho_C$, and the other is the concomitant $O$ distortion
in the $ab$ plane $\rho_\bot = \rho_O$.  Note that there is not rotational
degeneracy in the $ab$ plane due to the second-order anisotropy terms:
A specific direction is favored.

In addition to the wide ranging estimates for the critical exponent
$\beta= 0.25 - 0.42$ determined from the elastic properties discussed above,
there have been disparate results from other techniques.  Analysis of neutron
diffraction data for $\beta$ and $\nu$ indicated a simple mean-field transition
at $T_c$,\cite{nishi,pouget}
in agreement with one recent study of the specific heat.\cite{kuo}
In contrast, an earlier specific-heat measurement yielded an estimate
$\alpha \simeq 0.4$,\cite{sahling} whereas the most recently published result
claims $-0.15 \leq \alpha \leq 0.15$.\cite{liu}  The above results for
$\beta$ may also be compared with those from X-ray
scattering data:\cite{harris,lum} $\beta=0.26(3)$ and $\beta=0.37(3)$.
One may conclude from all of these results
that the most recent of studies of the critical behavior appear to favor a
scenario where the spin-Peierls transition in $CuGeO_3$ belongs to
one of the standard universality classes (Ising, XY, or Heisenberg).
It is of interest to note that some of the above results suggest
tricitical behavior ($\alpha = \frac12, \beta=\frac14$).  Simulations
of simple microscopic models of structural phase transitions indeed indicate
that the fourth-order coefficient of a corresponding Landau-type expansion
can be rather small.\cite{giddy}  Analysis of recent X-ray data on
$CuGeO_3$ demonstrates
that the critical region is rather small and that tricritical behavior is
evident if exponents are estimated using larger values of the reduced
temperature.\cite{lum}  Similar conclusions were recently made based on
Monte Carlo simulations of a magnetic system that was previously thought to be
tricritical.\cite{plum3}

\acknowledgements
I thank A Caill\'e for suggesting this problem, as well as M. Poirier,
M. Ain, B. Gaulin and M. Azzouz for useful discussions.
This work was supported by NSERC of Canada and FCAR du Qu\'ebec.
%==============================================================================
%  REFERENCES
%

%=============================================================================
%FIGURES
\begin{figure}
\caption{Schematic showing displacements of $Cu$ (open circles) and $O$
(filled circles) ions in the $ab$ (top planel) and $cb$ (bottom planel)
planes below $T_c$, after Ref. 8.  Not shown are the two $O$ and two $Ge$
ions per high-T unit cell (rectangles) which are undisplaced.}
\label{fig1}
\end{figure}

%==============================================================================

\begin{references}
%
\bibitem{mag} For recent reviews of magnetic properties, see e.g.,
S.B. Oseroff {\it et al.}, Phys. Rev. Lett. {\bf 74}, 1450 (1995);
O. Fujita {\it et al.}, {\it ibid} 1677 (1995); V. Kiryukhin and B. Keimer,
Phys. Rev. B {\bf 52}, R704 (1995).
%
\bibitem{adams} D.M. Adams, J. Haines, and S. Leonard, J. Phys. {\bf 3},
5183 (1991).
%
\bibitem{nishi} M. Nishi, J. Phys.: Cond. Mat. {\bf 6}, L19 (1994).
%
\bibitem{lorenzo} J.E. Lorenzo {\it et al.}, Phys. Rev. B {\bf 50}, 1278
(1994).
%
\bibitem{arai} M. Arai {\it et al.}, J. Phys. Soc. Japan {\bf 63}, 1661 (1994).
%
\bibitem{pouget} J.P. Pouget {\it et al.}, Phys. Rev. Lett. {\bf 72}, 4037
(1994).
%
\bibitem{kamimura} O. Kamimura {\it et al.}, J. Phys. Soc. Japan {\bf 63}, 2467
(1994).
%
\bibitem{hirota} K. Hirota {\it et al.}, Phys. Rev. Lett. {\bf 73}, 736
(1994).
%
\bibitem{roessli} B. Roessli {\it et al.}, J. Phys. {\bf 6}, 8469 (1994).
%
\bibitem{harris} Q.J. Harris {\it et al.}, Phys. Rev. B {\bf 50}, 12606 (1994).
%
\bibitem{stpaul} M. Saint-Paul, P. Monceau, and A. Revcolevschi, Solid State
Commun. {\bf 93}, 7 (1995).
%
\bibitem{nishi2} M. Nishi, O. Fujita, and J. Akimitsu, Physica B {\bf 210},
149 (1995).
%
\bibitem{chen} C.H. Chen and S-W. Cheong, Phys. Rev. B {\bf 51}, 6777 (1995).
%
\bibitem{poirier1} M. Poirier, M. Castonguay, A. Revcolevschi, and G. Dhalenne,
Phys. Rev. B {\bf 51}, 6147 (1995).
%
\bibitem{poirier2} M. Poirier, M. Castonguay, A. Revcolevschi, and G. Dhalenne,
unpublished.
%
\bibitem{poirier3} M. Poirier {\it et al.}, unpublished.
%
\bibitem{wink} H. Winklemann {\it et al.}, Phys. Rev. B {\bf 51}, 12 884
(1995).
%
\bibitem{yama} H. Yamaguchi, M. Yamaguchi, and T. Yagi, J. Phys. Soc. Japan
{\bf 64}, 1055 (1995).
%
\bibitem{landau} J.C. Toledano and P. Toledano,
{\it The Landau Theory of Phase Transitions} (World Scientific, Singapore,
1987); M.L. Plumer {\it et al.} in {\it Magnetic Systems with
Competing Interactions}, edited by H.T. Diep (World Scientific,
Singapore, 1994).
%
\bibitem{plum} See, e.g.,
M.L. Plumer and D.J.W. Geldart, J. Phys. C {\bf 16}, 677 (1983).
%
\bibitem{plum2} See, e.g.,
M.L. Plumer and A. Caill\'e, Phys. Rev. B {\bf 37}, 7712 (1988).
%
\bibitem{private} M. Poirier (private communication).
%
\bibitem{kuo} Y.-K. Kuo, E. Figueroa, and J.W. Brill, Solid State
Commun. {\bf 94}, 385 (1995).
%
\bibitem{sahling} S. Sahling, J.C. Lasjaunias, P. Monceau, A. Revcolevschi,
Solid State Commun. {\bf 92}, 423 (1994).
%
\bibitem{liu} X. Liu, J. Wosnitza, H. V. L\"ohneysen, and R.K. Kremer,
Phys. Rev. Lett. {\bf 75}, 771(C) (1995).
%
\bibitem{lum} M. Lumsden, H. Dabkowska, and B. D. Gaulin (unpublished).
%
\bibitem{giddy} A.P. Giddy, M.T. Dove, and V. Heine, J. Phys.: Cond. Mat.
{\bf 1}, 8327 (1989); S. Radescu, I. Etxebarria, and J.M. P\'erez-Mato,
{\it ibid} {\bf 7}, 585 (1995).
%
\bibitem{plum3} M.L. Plumer {\it et al.}, Phys. Rev. B {\bf 47}, 14312 (1993).

\end{references}
\end{document}